\documentclass[11pt,a4paper]{article} 
\usepackage{vmargin,color} 
\setmarginsrb{2.0cm}{2.7cm}{2.0cm}{1.0cm}{0.0cm}{0.0cm}{0.0cm}{1.0cm} 
\usepackage[french,english]{babel} 
\usepackage[T1]{fontenc} 
\usepackage{graphicx,float,setspace,amssymb,amsmath,cite}

\newcommand{\equa}[1]{\begin{eqnarray} \label{#1}} 
\newcommand{\auqe}{\end{eqnarray}} 
\newcommand{\tab}[1]{\begin{tabular}{#1}} 
\newcommand{\bat}{\end{tabular} \\ } 
 
\newcommand{\saut}{\vskip 0.5cm \noindent} 

\providecommand{\abs}[1]{\left\vert#1\right\vert} 
 
\begin{document} 
\selectlanguage{english} 
\title 
{{\bf
Spherical magnetic nanoparticles: magnetic structure and interparticle
interaction 
}} 
\author 
{ V. Russier 
\footnote{corresponding author russier@icmpe.cnrs.fr}\\ 
ICMPE, UMR 7182 CNRS and Universit\'e Paris Est, \\
2-8 Rue Henri Dunant, F-94320 Thiais, France.
\\ 
} 
\date{} 
\maketitle 
\thispagestyle{empty} 
\begin{singlespace} 
\abstract 
{ 
The interaction between spherical magnetic nanoparticles is investigated from 
micromagnetic simulations and ananlysed in terms of the leading dipolar interaction
energy between magnetic dipoles. We focus mainly on the case where the particles
present a vortex structure. In a first step the local magnetic structure in the 
isolated particle is revisited. For particles bearing a uniaxial magnetocrystaline 
anisotropy, it is shown that the vortex core orientation relative to the easy axis
depends on both the particle size and the anisotropy constant. When the particles
magnetization present a vortex structure, it is shown that the polarization of
the particles by the dipolar field of the other one must be taken into account in
the interaction. An analytic form is deduced for the interaction which involves
the vortex core magnetization and the magnetic susceptibility which are obtained 
from the magnetic properties of the isolated particle.     
\saut
} 
\end{singlespace} 
\eject
\section{ Introduction }
\label{intro} 
With the increasing progress in the synthethis of magnetic objects of 
nanometric scale such as spherical nanoparticles, nanodots, nanorings or 
layered films the diversity of systems made of such nano-objects as building
blocks either as 2D or 3D assemblies in non magnetic environment 
\cite{rev_skomski,martin,rev_bader,monolayer} or in colloidal suspensions 
as ferrofluids \cite{fer_fl_1,fer_fl_2} is continously growing. The magnetic 
behavior of magnetic nanometric particles either isolated or in nanostructured 
bulk materials is now quite well undertood both from experiments or numerical 
calculations \cite{rev_skomski,rev_bader,ohandley} but a precise knowledge of 
interparticles interactions and of their influence on the macrocopic properties 
is still needed. Indeed, the interparticle coupling has been investigated in a 
variety of systems, such as nanograins \cite{inter_grain_1,inter_grain_2,
inter_grain_3,inter_grain_4,spring_grain} nanorings \cite{inter_nanoring_1,
inter_nanoring_2,inter_nanoring_3} or cylindrical nanodots \cite{sd_inter_dot_1,
inter_dot_1,inter_dot_2,layer_nanodot} with a predominant attention paid on 
short range effects, such as exchange coupling, or on the influence of the 
coupling between single domain particles on the global magnetic properties. 
Conversely the long ranged interaction still deserves attention especially 
in cases where the magnetic structure of the isolated particle is complex 
(vortex \cite{bertotti,vortex_1,vortex_2,vortex_3,sph_nano1} or onion 
\cite{onion_1,onion_2} states for examples). A lot of work remains to be 
done on this point especially for spherical particles; in particular it 
seems important to develop models including the long ranged and anisotropic 
dipolar interaction. In the simple case of single domain particles the leading 
term in the interaction is the long range dipolar interaction which may lead to 
complex structures according to the shape of the particles on the one hand and 
the density and the dimensionality of the whole sytstem on the other hand 
\cite{dip_struc_2d1,dip_struc_2d2,dip_struc_2d3,dip_struc_3d}. In the case 
of particles with a non trivial internal magnetic structure, the interaction 
between particles is to be determined first. Indeed, it is generally admitted 
that when particles present a vortex structure, the resulting strong reduction 
of the magnetic moment at zero external field makes the dipolar interaction 
negligible. One aim of this work is to examine this point more precisely. In 
this work we focus on the interaction between spherical particles made of soft 
magnetic material (permalloy as an example) when they reach the vortex regime. 
We consider the simple situation of only two approaching spheres in a dumbell 
configuration. We especially compare the caculated interaction to the expected 
dipolar term. It is shown that two parameters characterizing the isolated 
particle play a central role: the magnetization of the vortex core and the 
suceptibility from which the polarization energy of one sphere in the dipolar 
field due the second one is calculated.

\section{ Magnetization structure and hysteresis } 
\label{part_1} 
Since our purpose is to model soft magnetic particles in a general way rather than 
to focus on particles of a given material, the magnetic characteristics are 
somewhat arbitrary and correspond  roughly to permalloy: the value of the exchange 
constant is $A_{x} = 1.10^{-11}J/m$, the saturation magnetization $J_s = 1T$ and 
the anisotropy is of uniaxial symmetry with a constant $K_1$ ranging from $K_1 = 0$
to $K_1 = 7.10^4 J/m^3$. The particle radius is varied from $R$ = 10 $nm$ to 45 
$nm$. In the following the particle volume will be denoted by $v_s$. We determine 
the magnetization structure in the framework of the micromagnetism formalism from 
the minimization of the total energy which is given by
\equa{ener_1_spere}
E_{tot} &=& E_{x} + E_{a} + E_{dm} + E_{Z}                       \nonumber \\
  &=& \int_\Omega \left[ A_x \Sigma_i (\nabla m_i(\vec{r}))^2
  + K_1(1 - (\vec{m}(\vec{r}).\hat{a})^2)
- \frac{1}{2} \mu_0 M_s \vec{m}(\vec{r}) . \vec{H}_{dem} 
- \mu_0 M_s \vec{H}_{ex} . \vec{m}(\vec{r}) \right]  
d \vec{r} 
\auqe
where $E_{x}$, $E_{a}$, $E_{dm}$ and $E_{Z}$ are the exchange, anisotropy,
demagnetizing and Zeeman terms respectively, $\hat{a}$ is the unit vector in
the direction of the easy axis and $\Omega$ is the total volume of magnetic 
material, wich can include more than one particle. $\vec{m}(\vec{r})$ is the 
reduced magnetization density, related to the total magnetization by 
$M_t = \int_{\Omega} M_s\vec{m}(\vec{r}) d\vec{r}$. The calculations are 
performed with the micromagnetic code MAGPAR \cite{magpar} which is based upon 
a finite element method. The problem includes two length scales, namely the 
exchange length, $l_{ex} = (2\mu_0 A_x/J_s^2)^{1/2}$ and the Bloch wall length
$l_{B}$ = $\sqrt{A_x/K_1}$. Here, given the parameters chosen we have
$l_{ex} = 5.013 nm$ and $l_{B} > 12 nm$. The value of the dimensionless 
parameter, $K = 2K_1/(\mu_0 M_s^2)$ as defined in \cite{bertotti} ranges in 
between 0 and 0.175. The mesh used in the calculations is such that the largest 
tetaedron size is smaller than $l_{ex} $ which imposes typically a mesh with 
$N_{fe} \sim 10^5$ elements for one sphere of radius $R \simeq $ 40 $nm$. We 
first calculate the magnetic structure of one isolated sphere in terms of both 
the size and the anisotropy constant $K_1$. In order to characterize the 
magnetic state in the particle, we consider the local magnetization profile, 
$\vec{m}({\bf r})$, which in the vortex regime, is decomposed in its cylindrical 
components using the vortex axis, say $\hat{v}$, as the cylindrical axis
\equa{proj}
\vec{m}({\bf r}) = m_v \hat{v} + m_{\varphi} \hat{\varphi} + m_{\rho}\hat{\rho}
\auqe
where $\hat{\rho}$ and $\hat{\varphi}$ are the radial and tangential unit vectors 
of the projection of {\bf r} in the plane normal to $\hat{v}$. In the following, 
hatted letters denote unit vectors. 
The axis $\hat{v}$ is defined and actually determined as the mean direction
of the local magnetization in the central part of the vortex, as shown in figure
(\ref{def_struct_vortex}).
In a first step we focus on the behavior of 
the magnetization $M$ in terms of the external field, $H_{ex}$, especially for the 
variation of the field from small values up to saturation field; however we do not 
focus on the nucleation field. First of all, as is well known, small particles
up to a threshold value, $R_{SD}$, are uniformly magnetized in a single magnetic 
domain and the hysteresis curve is a square. With our set of parameters, we get 
$R_{SD}$ = 18 $nm$, for $K_1$ = 0 and 22 $nm$ for $K_1$ = 3 $10^4$ $J/m^3$, in 
agreement with the result of the micromagnetic calculations of \cite{calc_sph_c} 
and with the estimation given in \cite{bertotti}. Then a vortex structure is 
obtained, characterized by a vanishing value of the radial component 
$m_{\rho}({\bf r})$, and $\abs{m_{\varphi}(\rho)}$ varying from 
$\abs{m_{\varphi}(\rho)} = 0$ inside the vortex core, $\rho < r_c$, to 
$\abs{m_{\varphi}(\rho)} = 1$ in the vicinity of the particle surface 
$\rho \sim$ $\rho_{max} = R sin(\theta(z))$. At zero external field, the vortex 
direction, $\hat{v}$ is arbitrary when $K_1$ = 0, while for $K_1 \neq 0$, the 
direction taken by $\hat{v}$ relative to the easy axis $\hat{a}$ is controlled by 
the anisotropy energy which tends to allign $\vec{m}(\vec{r})$ on $\hat{a}$. 
The anisotropy energy depends on both the value of $K_1$ and the volume fraction 
of the particle where $\hat{m}(\vec{r})$ is oriented parallel or antiparallel to 
$\hat{a}$ ($\abs{\hat{m}.\hat{a}} \simeq 1$). The ratio of the volume fraction 
corresponding to the vortex core $v_c$, characterized by $\hat{m}$ oriented 
parallel to $\hat{v}$, to the volume fraction where $\vec{m}(\vec{r})$ is oriented 
normal to $\hat{v}$ is directly related to the volume of the particle, $v_{s}$. 
Roughly speaking the ratio of the total volume where 
$\abs{ \hat{m}.\hat{a}} \simeq 1$ is either $v_c/v_s$ or $(1/2)(v_s - v_c)/v_s$
if $\hat{v}$ is parallel or normal to $\hat{a}$ respectively. According to this
scheme the stability condition for the vortex direction $\hat{v}$ to be normal 
to $\hat{a}$ reads $v_c < (1/3)v_s$. We can refine this very crude determination 
of the thershold by introducing the magnetization profile and imposing in 
(\ref{proj}) $m_{\rho} = 0$. Then $m_\varphi = \sqrt{1 - m_v^2}$ and we get
\equa{anis_v_a}
E_a^{//} = 
- K_1 \int_{-R}^{R}dz \int_{0}^{R(z)} m_v(\rho)^2
2\pi \rho d\rho   \nonumber \\
E_a^{\perp} = 
- K_1 \int_{-R}^{R}dz \int_{0}^{R(z)}\frac{1}{2}(1 - m_v(\rho)^2) 
2\pi \rho d\rho 
\auqe
for $\hat{v}$ parallel or normal to $\hat{a}$ respectively. Then we assume that 
the component $m_v(\rho)$ depends on $\rho$ only through $r^*$ = $\rho/r_{sc}$, 
$r_{sc}$ being the pertinent scaling length (either $l_{ex}$ for $K_1$ = 0 or a 
function of both $l_{ex}$ and $l_{B}$ otherwise) and we neglect its dependence 
with respect to $z$. Hence, exploiting $m_v(\rho>r_c) = 0$, we set the upper 
bound in the integral over $m_v$ to $\infty$ and we write 
$(E_a^{\perp} - E_a^{//})$ in the form 
\equa{anis_v_a2}
E_a^{\perp} - E_a^{//} = 
- K_1 \left( \frac{1}{2} v_s - 3 R r_{sc}^2 I \right)
\makebox[2.7 cm]{with} I = \int_{0}^{\infty} m_v(r^*)^2 2\pi r^*dr^*
\auqe
The stability condition for a vortex normal to $\hat{a}$ is now
\equa{stab_v_a}
\frac{R}{r_{sc}} > \sqrt{\frac{9I}{2 \pi}}
\auqe
which must be read as $R > r_{sc} \sqrt{9I/(2 \pi)}$ = $R_{th}(K_1)$ when 
$R$ is varied at constant $K_1$, or conversely as $r_{sc} < R\sqrt{2 \pi/(9I)}$
when the role of the magnetic characteristics is investigated for a given 
particle size. Equation (\ref{stab_v_a}) can be rewritten in a more convenient
form for the practical calculatio$J/m^3$ns of $I$, namely : $S^*$ =
$\sqrt{9I^*(r_{sc})/(2\pi)} < 1$ where $I^*$ is given by (\ref{anis_v_a2}) with 
$r^*$ replaced by $(r/R)$ and the upper bound replaced by $(r/R)_{max}$ = 1.
Of course $I^*$ is then dependent on the value of $r_{sc}$ which is emphasized
by the notation $I^*(r_{sc})$. In any case, such an estimation is not supposed 
to provide an accurate determination of the threshold value of either $R$ or 
$r_{sc}$ for the orientation of $\hat{v}$ normal to $\hat{a}$ but to predict at a
qualitative level the effect of either the particle size or the magnetic 
parameters on the direction taken by the vortex. We can scale the vortex radius,
$r_c$ on the smallest of the two characteristic lengths, $r_{sc}$ = 
$inf(l_{ex}, l_{B})$; however, this scaling may be taken with care and instead 
we can consider, when $l_B$ increases, a scaling radius in the form of a 
function $r_{sc}(l_{ex}, l_{B})$. Notice that when using $S^*$ instead of $S$
one has not to explicit the dependence of $r_{sc}$. Therefore, at both $A_x$ and 
$J_s$ kept constant, we deduce from equ. (\ref{stab_v_a}) that $\hat{v}$ gets 
$\perp$ $\hat{a}$ when the sphere radius is increased at $K_1$ constant or when 
$K_1$ is increased at $R$ constant. As we shall see in the following the 
stability condition for the orientation of $\hat{v}$ relative to $\hat{a}$ 
agrees with this qualitative conclusion. Notice that the orientation of the 
vortex relative to the axis of easy magnetization is also found to be size 
dependent in the case of the cubic anisotropy \cite{calc_sph_c} : in this latter 
case, the vortex is parallel to the axis of easy magnetization in large spheres.

The orientation of the vortex relative to $\hat{a}$ can be determined from the 
magnetization profile, $\hat{m}(\vec{r})$ as well as from 
the magnetization curve in terms of the external field, $M(H_{ex})$ by chosing
the direction of the external field, $\hat{h}_{ex}$ either
parallel or normal to $\hat{a}$. Indeed, we expect the magnetization process
to differ according to the direction of the external field relative to 
the vortex one. We keep in mind the well known behavior of the vortex in
the flat cylindrical nanodots where the magnetization is found to result
from the shift of the vortex core when the field is applied normal
to the vortex direction. Here, in the case of $\hat{v}(H_{ex}=0)$ $\perp$ 
$\hat{a}$ we expect a similar behavior for small values of the external 
field when $\hat{h}_{ex}$ = $\hat{a}$, up to the rotation of the vortex core 
along the direction of the field for high values of $H_{ex}$ before the 
magnetization in the whole volume of the sphere becomes oriented along 
$\hat{h}_{ex}$. Morever, in this case, we expect to have no remanence in the 
direction of the field, since in the vicinity of $H_{ex}=0$ there is no net 
magnetization normal to the vortex direction. 
On the other hand a non vanishing magnetization at $H_{ex} = 0$, corresponding
to the vortex polarization in the direction $\hat{v}$ $\perp$ $\hat{a}$ = 
$\hat{h}_{ex}$ will be obtained. Conversely, the magnetization curve 
corresponding to $\hat{h}_{ex}$ $\perp$ $\hat{a}$ still for a sphere 
characterized by $\hat{v}(H_{ex} = 0)$ $\perp \hat{a}$ will present the more
usual shape of a loop located in between $\pm H_c$ with a non zero remanence 
corresponding to the vortex polarization.

Then we focus on the external field induced magnetization in the spherical 
particle. As is generally obtained in nanodots or spherical soft magnetic particles 
\cite{nanodot,calc_sph_1,calc_sph_2}, the magnetization $M$ in the direction of 
the external field is found to vary nearly linearly with respect to the field, at 
least in the vicinity of $H_{ex}$ = 0 and of course away from switching points 
where the vortex reverses as a whole. Such a linear behavior is observed both when 
$\hat{h}_{ex}$ = $\hat{v}$ or $\hat{h}_{ex}$ $\perp$ $\hat{v}$. (or equivalently 
$\hat{h}_{ex}$ $\perp$ $\hat{a}$ or $\hat{h}_{ex}$ = $\hat{a}$ when 
$R > R_{th}(K_1)$). This means that the susceptibility $\chi$ defined as 
\equa{chi}
\frac{\partial M}{\partial H_{ex}} = \chi
\auqe
does not depend on the value of the field to a very good approximation. We 
emphasize that the suceptibility is well defined for particles in the vortex
regime since no multidomain state occurs and therefore a demagnetized state
at $H_{ex} = 0$ can be ruled out. Notice that the value of $\chi$ depends on 
the direction of the field as will be discussed below, and we should 
distinguish $\chi_{\parallel}$ from $\chi_{\perp}$ according to the direction 
of the field relative to $\hat{v}$. When such a distinction is not necessary 
it will be omitted to lighten the notations and $\chi$ is to be understood as 
its value corresponding to the orientation chosen for the field. Only in the 
case of an external field direction $\hat{h}_{ex}$ neither parallel nor normal 
to the vortex direction the consideration of the two values of $\chi$ is 
necessary. The independence of $\chi$ with respect to $H_{ex}$ can be exploited 
for obtaining the variation of the total energy with respect to the external 
field. We consider the variation of $M$ starting from $H_{ex}$ = 0 to a value of 
$H_{ex}$ such that no switching of the magnetization occurs up to $H_{ex}$ and 
we analyse the corresponding variation of the magnetization, $\Delta M$ as the 
polarization of the sphere induced by the field. We have $\Delta M(H_{ex})$ 
= $\chi H_{ex}$. On the other hand, we can deduce $\Delta M(H_{ex})$ from 
the energy, $E(H_{ex})$ by writting an equilibrium equation
\equa{pol_ener1}
\frac{\partial E_{tot}(\Delta M)}{\partial \Delta M} = 0
\auqe
which determine the equilibrium value of $\Delta M$. The
total energy depends explicitly on $H_{ex}$ through the Zeeman term, 
$-\mu_0H_{ex}(m(0)\hat{v}.\hat{h}_{ex} + \Delta M)$, where we have expressed 
the permanent magnetization in the absence of the field as 
$\vec{M}(H_{ex} = 0)$ = $m(0)\hat{v}$, $m(0)$ being the magnitude of 
the vortex core magnetization in the absence of the field. Then from 
(\ref{pol_ener1}) we get
\equa{pol_ener2}
\frac{\partial}{\partial \Delta M}\left( E_{dm} + E_{x} + E_{a} \right ) =
\mu_0 H_{ex} 
\auqe
Therefore we get the variation of the total energy in the form 
\equa{pol_ener3}
E = E(H_{ex} = 0) + \int_{0}^{\Delta M} \mu_0 H_{ex}(\Delta M') d\Delta M'
- \mu_0 H_{ex}(m(0)\hat{v}.\hat{h}_{ex} + \Delta M)  \label{pol_ener3a} \\
= E(H_{ex} = 0) + \mu_0 \frac{\Delta M^2}{2 \chi}
- \mu_0 H_{ex}(m(0)\hat{v}.\hat{h}_{ex} + \Delta M) \label{pol_ener3b}
\auqe
where we have used $\Delta M(H_{ex})$ = $\chi H_{ex}$. 
Notice that both (\ref{pol_ener2}) and (\ref{pol_ener3a}) are exact equations
while (\ref{pol_ener3b}) holds only in the case of a linear dependence of
$\Delta M(H_{ex})$ with respect to $H_{ex}$.
The second term of the r.h.s. of (\ref{pol_ener3a}) or (\ref{pol_ener3b}) has 
a simple interpretation: it is the energy of polarization of the sphere and 
corresponds to the energy cost of the reorientation of the magnetization 
inside the sphere. Equ. (\ref{pol_ener3b}) is to be compared to the expression 
of the energy density of an array of coupled dots presenting a vortex structure 
obtained in \cite{dot_array}; more precisely the polarization energy coincides 
with the second term of equ.(5) of Ref. \cite{dot_array} where the induced
magnetization in the dot is related to the vortex shift, $s$. The polarization
energy can be written equivalently as $\mu_0(\Delta M H_{ex})/2$ where $\Delta M$ 
is to be understood as the induced moment due to the external field. Finaly, 
$\chi$ can be related to the variation of the energy minus the Zeeman term
\equa{chi_bis}
\chi = \frac{1}{\mu_0 H_{ex}}\frac{\partial (E_{tot} - E_Z) }{\partial H_{ex}}
\auqe

\section{ Interaction energy between magnetic spheres } 
\label{part_2} 
Now we focus on the determination of the interaction energy between two magnetic
nanoparticles in terms of the interparticle distance, $r_{12}$. The interaction 
energy is defined in a usual way 
\equa{delta_e}
E_{int}(1,2) = E_{tot}(1,2) - E_{tot}(r_{12} \rightarrow \infty) 
\auqe
where $E_{tot}$ denotes the total energy of the two particles system, and 
$(1,2)$ is a short notation for the orientation and location variables of the
particles when they are brought together. We expect to get a form dictated by 
the dipolar interaction between the magnetic moment of the approaching spheres 
which reads
\equa{inter_dip}
E_{dip} = 
\frac{\mu_0 m_1 m_2}{4 \pi r_{12}^3} d_{112}(\hat{m}_1, \hat{m}_2, \hat{r}_{12})
\\ \label{d_112}
d_{112}(\hat{m}_1, \hat{m}_2, \hat{r}_{12}) = 
\hat{m}_1.\hat{m}_2 - 3(\hat{m}_1.\hat{r}_{12})(\hat{m}_2.\hat{r}_{12})
\auqe
where $m_i$ are the magnitude of the magnetic moments, $\hat{m_i}$ and 
$\hat{r}_{12}$ the unit vectors in the direction of both moments and of the 
vector joigning the two particles and $d_{112}$ is the angular function 
characteristic of the dipolar intercation. In the case of single domain
particles $m_i = M_s v_s$ where $M_s$ is the saturation 
magnetization of the particles, and the orientations $\hat{m}_i$ result from 
the minimum of $d_{112}$ given in (\ref{d_112}). For particles without 
magnetocristalline anisotropy, this gives obviously : $\hat{m}_1$ = $\hat{m}_2$
= $\hat{r}_{12}$ and $d_{112}$ = -2. On the other hand, if the 
magnetocristalline energy is non zero on both particles with easy axes 
$\hat{a}_i$, the orientations $\hat{m}_i$ will result of the interplay 
between the anisotropy energy tending to align $\hat{m}_i$ on $\hat{a}_i$ 
and the energy (\ref{inter_dip}) tending to minimize the angular function 
(\ref{d_112}). Furthermore if $K_1$ takes a non vanishing value only 
in one particle say $i = 1$ and if this value is large enough to impose 
$\hat{m}_1$ = $\hat{a}_1$, $\hat{m}_2$ must orient in the dipolar field due 
to particle $1$ {\it i.e.} in such a way that $d_{112}$ =
$\hat{m}_2.\hat{a}_1 - 3(\hat{m}_2.\hat{r}_{12})(\hat{a}_1.\hat{r}_{12})$
is minimum. 

Now we consider the case of particles large enough to present a vortex 
structure. In this case, the orientations of the effective moments of the 
particles are the vortex directions, $\hat{v}_i$, and the values of the
moments are no more equal to $M_s v_s$ but correspond to the vortex cores 
magnetizations and can be obtained from the magnetization curves $M(H_{ex})$. 
Let us introduce the coefficients $\alpha_i$ = $m_i/(M_s v_s)$
(in the following, we shall only consider the case of identical particles, so 
we drop the index $i$). $\alpha$ depends of course on the location of the
second particle, which will be denoted in short by $\alpha(r_{12},d_{112})$
or $\alpha(1, 2)$.
The value taken by $\alpha$ is not trivial since on the one hand it must be
determined from the characteristics of the isolated particle and from 
the polarization of the particle by the dipolar field of the second one. 
A simple approximation for the interaction energy can 
be built in the framework of the dipolar approximation by considering that each
particle is in the dipolar field of the other one. Then, we have to take into 
account two contributions. The first one which corresponds to (\ref{inter_dip}),
is nothing but $-m_1 H_{dip}(r_{12}) (\hat{v}_1.\hat{h}_{dip}(2,1))$ where
$H_{dip}(r_{12}) \hat{h}_{dip}(2,1)$ is the dipolar field created at ${\bf r}_1$
by particle at ${\bf r}_2$ and the second one is twice the polarization energy 
of each sphere in the field of the second one. The second contribution has been
introduced in (\ref{pol_ener3b}) for one particle in a constant external field.
In the present case, the role of $m(o)$ is played by $M_s v_s \alpha(\infty)$
while the induced moment in the direction of the dipolar field is 
\equa{ind_mom1}
\vec{p} = p\hat{h}_{dip} = \chi H_{dip}(r_{12})\hat{h}_{dip}
\auqe
We first consider the case where the vortex $\hat{v}_i$ is free to orient in
the direction of the dipolar field due to particle {\it j $\ne$ i}. This is the
most general case since it corresponds to both the absence of anisotropy or 
particle large enough for the vortex to be normal to the easy axis. In this
case we have
\equa{ind_mom2}
p\hat{h}_{dip} = 
(\alpha(r_{12}, d_{112}) - \alpha(\infty))M_s v_s \hat{h}_{dip} 
= 
\Delta\alpha(r_{12}, d_{112})M_s v_s \hat{v}  
\auqe
Now adding twice the second term of (\ref{pol_ener3b}) to the total dipolar
energy we get for the interaction energy
\equa{inter_ener}
E_{int}(1,2) = 
\frac{\mu_0 (M_s v_s)^2}{4 \pi r_{12}^3}
\alpha(\infty)(\alpha(\infty) + \Delta\alpha(r_{12}, d_{112}))
d_{112}(\hat{m}_1, \hat{m}_2, \hat{r}_{12})
\auqe
which coincides with the interaction energy between polar polarizable hard 
spheres \cite{polar}. This is the important result of this section. It relates 
the interaction energy to the magnetic charateristics of the isolated spheres,
namely, $\alpha(\infty)$ and $\chi$ through $\Delta \alpha$. Notice that this
form for the interaction energy should hold not only in the case of two 
particles but also more generaly for an assembly of particles. In the latter 
case, the solvation of the total dipolar field and thus the determination of 
$\Delta\alpha(r_{12}, d_{112})$ becomes a difficult task. In the simple case 
of two particles, introducing $u = \chi/(4\pi R^3)$ (= $\chi^*/3$ where 
$\chi^*$ = $\chi/v_s$ is the reduced susceptibility) we get
\equa{chi2}
\Delta\alpha = \frac{-u\alpha(\infty)d_{112}}{((r_{12}/R)^3 + ud_{112})} 
\auqe
\section{ Results }
\label{results}

We analyse first the magnetic behavior of the isolated particle with a special 
attention paid on the characterization of the vortex structure at low external 
fields. The magnetization curve is displayed in figure (\ref{hyst_r45_k0}) for 
$K_1 = 0$ and $R$ = 45 $nm$ and the corresponding magnetic structure, through 
the local magnetic moment components (\ref{proj}), is shown in figures 
\ref{prof_rem_r45_k0} and \ref{prof_coer_r45_k0} for the remanent state and in 
the vicinity of the coercive field, before and after the reversal of the vortex 
core. Since $K_1 = 0$, the vortex direction, $\hat{v}$ coincides with the 
direction of the external field. These results put in evidence the vortex 
structure and in particular the vortex core is reversed as a whole at the 
coercive field, with a nearly frozen $\vec{m}(\vec{r})$ structure. Moreover, 
we find that the reversal of $\vec{m}(\vec{r})$ results from a global rotation 
since the component $m_{\varphi}$ changes sign. When $K_1 \neq 0$ as described 
at the qualitative level in section (\ref{part_1}) $\hat{v}$ is parallel to the 
easy direction $\hat{a}$ for small values of $R$, and becomes normal to $\hat{a}$ 
beyond a $K_1$ dependent threshold value  $R_{th}(K_1)$. When 
$\hat{a} \perp \hat{v}$, the vortex core is free to rotate in the plane normal 
to $\hat{a}$ and therefore will orient parallel to the external field if 
$\hat{h}_{ex} \perp \hat{a}$. As an example, we show in figure 
(\ref{hyst_r45_k3e4}) the magnetization curve for the two directions of the 
external field $\hat{h}_{ex} = \hat{a}$ and $\hat{h}_{ex} \perp \hat{a}$, in the 
case $R$ = 45 $nm$. Moreover, in the former case, the magnetization parallel and 
normal to the external field, $M_{\parallel}$ and $M_{\perp}$ are displayed. The 
magnetization behavior in terms of the external field corresponds to the 
situation $\hat{v} \perp \hat{a}$; indeed, the remanence vanishes when 
$\hat{h}_{ex} \parallel \hat{a}$, while $M_{\perp}$ takes a nearly constant 
value when $H_{ex}$ is varied in the central part of the $M_{\perp}(H_{ex})$ curve. 
Moreover this value coincides with the remanence obtained for 
$\hat{h}_{ex} \perp \hat{a}$ or equivalently $\hat{h}_{ex} \parallel \hat{v}$ 
and therefore corresponds to the vortex core magnetization. The independence 
of $M_{\perp}$ with respect to $H_{ex}$ in the central part of the 
$M_{\perp}(H_{ex})$ curve shows that the variation of the magnetization 
$M_{\parallel}(H_{ex})$, i.e. in the direction normal to $\hat{v}$,
corresponds to a shift of the vortex core normal to the direction of the field. 
The vortex core magnetization is thus nearly constant and given by the
value of $M_{\perp}$ in that part of the curve.
This is in agreement with the magnetization process obtained in the 
flat nanodot vortex structures. Finally $M_{\perp}$ sharply vanishes when the 
vortex rotates in the direction of the field, where the magnetization curve
$M_{\parallel}$ whith $\hat{h}_{ex} = \hat{a}$ presents the hysteretic
wings, similar also to what is found in the flat nanodot case where however 
this last value of the field corresponds to the vortex anhihilation prior to
the saturation of the dot. The behavior of $M(H_{ex})$ outlined above is 
coroborated by the evolution with the value of the field of the structure of 
$\hat{m}(\vec{r})$, shown on figure (\ref{prof_shift_r45_k3e4}), where we see 
that the linear variation of $M_{\parallel}(H_{ex})$ in the central part of 
the curve can be associated to a shift of the vortex in a direction normal to 
$\hat{m}$. From the evolution of $M_{\parallel}$ for $\hat{h}_{ex} = \hat{a}$ 
with the particle size, displayed on figure (\ref{hyst_taille_k3e4}), 
one can determine the threshold value $R_{th}(K_1)$ beyond which $\hat{v}$ 
is normal to $\hat{a}$. Here we find $R_{th} \simeq$ 28 $nm$ for 
$K_1 = 3.10^4 J/m^3$ (strictly speaking, 26 $nm$ $\le R_{th} \le$ 30 $nm$).
We have calculated numerically the integral $I^*(r_{sc})$ defined after 
equation (\ref{anis_v_a2}) from which we find that the threshold condition
(\ref{stab_v_a}) is satisfied (see table \ref{num_stab_v_a}) in good agreement 
with the onset of the vortex structure deduced from the magnetization. 
Indeed, from this calculation, we get $S^*$ = 1 for $R$ = 26.5 $nm$ when 
$K_1$ = $3.10^4$ $J/m^3$ and thus $R_{th}$ = 26.5 $nm$ in agreement with the 
value deduced from the behavior of the magnetization $M(H_{ex})$. 
Similarly by decreasing $K_1$ at constant $R = 45 nm$, we find that the range
of external field where the vortex is normal to $\hat{a}$ is reduced and 
then vanishes for $K_1 = 2.10^3 J/m^3$. Therefore we confirm our prediction
that $\hat{v} \perp \hat{a}$ for $K_1 > K_{1 th}(R)$ at constant $R$. Then the 
value of $\chi$ is determined from the slope of the magnetization
curve, $M(H_{ex})$ in terms of $H_{ex}$. The results are listed in table 
\ref{tab_chi}. We also check that equ. (\ref{chi_bis}) is satisfied 
(see table \ref{tab_chi}). 
\saut
\saut
\subsection{Interaction beween particles}
\label{res_inter}
We first consider the case of monodomain particles; for this we chose 
$R$ = 10 $nm$. As expected the interaction energy is exactly given by the 
dipolar term with $m_i$ = $M_s v_s$. When $K_1$ = 0 for both particles, the 
energy minimization leads to $d_{112}(1, 2)$ = -2 and we thus mainly test the 
$1/r_{12}^3$ dependence of the interaction. On the other hand, we have also 
considered the case where only one particle bears a non vanishing uniaxial 
anisotropy with a value of $K_1$ large enough to impose the orientation of its 
moment, $\hat{m}$ parralel to the easy axis $\hat{a}$. Then the moment of the 
second particle orient itself in the field of the fixed particle in order to 
minimise the angular function $d_{112}$. This provide an additionnal test of 
the behavior of the interaction through its angular dependence. The results  
are displayed in table \ref{tab_d112}.
Now we consider the vortex regime with particles of radius $R$ = 35$nm$
or $R$ = 45 $nm$. We start from particles without anisotropy, $K_1$ = 0. In 
this case only one value for the susceptibility, $\chi_{\parallel}$, is to be 
considered, since the vortex allign spontaneously in the direction of the 
dipolar field. The two parameters involved in the expression of the interaction, 
$\chi$ and $\alpha(\infty)$, are determined first from the magnetization
curve of the isolated particle. As a first test, we look at the angular
dependence of the interaction energy. To this aim we start from the two
spheres at a large distance and we minimize the total energy corresponding
to non interacting spheres. Then we decrease the distance $r_{12}$ down to
a not too small value of the ratio $r_{12}/R$ and we perform a rotation of 
one sphere, say $2$, arround the other one which is kept fixed. In this
first calculation, we just calculate the components of the energy without 
minimization; we thus obtain the energy at a fixed value of local magnetic
structure in the spheres, disregarding the polarization energy. The result
is displayed in figure (\ref{en_rot_r35_k0}) in the case $r_{12}/R$ = 4 and 
different values of the angular function $d_{112}$ calculated by using 
$\hat{m}_i$ = $\hat{v}_i$. We clearly obtain a linear dependence of $E_{int}(1,2)$
in terms of $d_{112}(1,2)$, and moreover the proportionality factor is exactly 
the result of the dipolar interaction, as deduced from (\ref{inter_dip}). We 
thus conclude that when the structure inside the spheres is frozen, the resulting 
interaction energy is indeed given by the dipolar interaction between the vortex 
cores. Then we consider the interaction energy after relaxation of the structure
in the spheres, namely from the result of the total energy minimization in terms 
of the distance between particles. As expected and in agreement with eq. 
(\ref{inter_ener}) the value we get for $d_{112}(1,2)$ is very close to $d_{112}$ 
= -2 especially for short distances. The result for the interaction is shown in 
figure (\ref{en_r35_k0_d2}). We also compare the result corresponding to the 
dipolar interaction including the polarization energy or without this las term. 
This later approximation amounts to model the interaction by that between the 
dipoles corresponding to the isolated particles vortex cores. The approximation 
introduced in (\ref{inter_ener}) is in very good agreement with the calculated 
result, for distances down to $r_{12}/R$ $\sim$ 2.75, and the agreement for 
$r_{12}/R$ = 2.5 is still fairly good. Moreover we see that the inclusion of the 
polarization energy is quite important; indeed, the dipolar interaction 
calculated with the moments resulting from the isolated particles vortex cores 
reproduces the interaction only for distances larger than 3.35$R$. 

In the case of particles with non zero uniaxial anisotropy, we focus on 
a situation where the vortex direction, $\hat{v}$, is normal to the easy
axis at zero external field. As an example we choose $K_1 =$ 3.10$^4 J/m^3$
and either $R =$ 45 $nm$ or $R =$ 35 $nm$. One can impose the plane in 
which the vortex is free to rotate {\it via} the direction chosen for the 
easy axis. Here we consider two situations where the two particles have 
the same easy axis, say $\hat{a}$ = $\hat{z}$ and the unit vector joigning 
the particles $\hat{r}_{12}$ is either normal or parallel to $\hat{a}$. 
Thus the equilibrium configuration of the particles corresponds to $\hat{v}_1$ 
= $\hat{v}_2$ = $\hat{r}_{12}$ and $d_{112}(1,2)$ = -2 in the former case and 
$\hat{v}_1$ = - $\hat{v}_2$ $\perp$ $\hat{r}_{12}$ and $d_{112}(1,2)$ = -1 in the 
latter case. The results are summarized in figure (\ref{en_tot_ren}) where we plot 
the interaction energy normalized by the value at the shortest distance considered,
$r_{12}$ = 2.25$R$. The interaction energy is still very close to the dipolar plus 
polarization energy, eq (\ref{inter_ener}) when $d_{112} = -1$, while in the case 
where the vortices are in line, the agreement for short distances is more 
qualitative. This is mainly due to an underestimation of the induced polarization 
by the dipolar field. We are lead to this conclusion by fitting the values of the 
parameters $\alpha$ and $u$ in order to reproduce the calculated interaction energy
by equ. (\ref{inter_ener}). Doing this we can reproduce the calculated interaction 
energy only by using a non negligible enhancement of $u$ while the fitted value of 
$\alpha$ remains very close to that calculated on the isolated sphere. The fitted 
results are also displayed in figure (\ref{en_tot_ren}). To get a similar agreement
with what is obtained in the case $K_1$ = 0 with $d_{112}$ = $-2$, the fitted value
of $u$ and $\alpha$ are 1.25$u^{calc}$, 1.03$\alpha^{calc}$ and 1.45$u^{calc}$, 
1.05$\alpha^{calc}$ for $R$ = 35 $nm$ and 45 $nm$ respectively.  

The results of this work are twofold. First we have precised the local magnetic 
structure in the sphere, and shown that beyond the well documented single domain 
to vortex transition in the case of a uniaxial anisotropy the vortex direction is 
normal to the easy axis once the particle radius is larger than a threshold value, 
$R_{th}(K_1)$ for which a simple estimation is given. Then the interaction between 
particles is shown to present a dipolar character depending on two parameters 
characterising the isolated particle, namely the vortex core magnetization and the 
suceptibility. The vortex core magnetization is strongly reduced when compared to 
the saturation magnetization $M_s$ which is quantified by the parameter 
$\alpha$ $\sim$ 0.2 and this makes the interaction rather small but nevertheless 
non negligible. The order of magnitude of the interaction energy at distance 
$r_{12}$ = 2.25$R$ is slighltly smaller than the barrier necessary to reverse the 
vortex core. However due to both its long range and its anisotropy the dipolar 
interaction is likely to lead to measurable effects in experimental assemblies of 
such particles. On a qualitative point of view, we do think that some of the 
finding of \cite{sph_nano1} are in agreement with the manifestation of dipolar 
effects, namely the tendency to form chains and to allign the vortex cores. In ref 
(\cite{sph_nano1}) a micromagnetic simulation was already performed and was in 
agreement with the experiments; however, here we go a step forward by clearly 
pointing the dipolar character of the interaction between spherical nanoparticles. 
This allows us to predict that in a general way the behavior of dipolar and 
polarizable hard spheres will be transferable to assemblies of such particles even 
in the vortex regime. In this field, a very rich panel of structures is expected 
for both 2D systems \cite{dip_struc_2d1} (and reference therein), 
\cite{dip_struc_2d2,dip_struc_2d3} and 3D systems \cite{dip_struc_3d}.
\section*{Acknowledgements}
\label{acknow}
The author acknowledges stimulating and important discussions with 
Dr. Y. Champion (ICMPE, CNRS, Thiais France), 
Dr. F. Mazaleyrat (SATIE, ENS Cachan France)
and Pr. L. Bessais (ICMPE, CNRS, Thiais France).
\eject
\begin{singlespace} 
 
\end{singlespace} 
\eject 
%
\begin{table}[H] \begin{doublespace}
\caption{ {\label{num_stab_v_a}} 
Value of $S^*$ = $\sqrt{9I^*(r_{sc}/(2\pi)}$ involved in the stability condition
(\ref{stab_v_a}). $K_1$ = 3. $10^4 J/m^3$.
$I^*$ is defined by equ.~(\ref{anis_v_a2})  with $r_{sc}$ = $R$.
}
\vskip 1.0 cm
\tab{| l | *{4}{l} |}
\hline
$R$ ($nm$) & 26 & 30 & 37 & 45 \\
$S^*$    & 1.042 & 0.685 & 0.596 & 0.533 \\
\hline
\bat \\
\end{doublespace} \end{table}
%
\begin{table}[H] \begin{doublespace}
\caption{ {\label{tab_chi}}  
Reduced magnetic susceptibility calculated from $(a)$ : equation~(\ref{chi});
$(b)$ equation~(\ref{chi_bis}).
}
\vskip 1.0 cm
\tab{| *{6}{l} |}
\hline
$K_1$ & $R$ & $\chi_{\parallel}^*$ $^{(a)}$ & $\chi_{\parallel}^*$ $^{(b)}$ &
$\chi_{\perp}^*$ $^{(a)}$ & $\chi_{\perp}^*$ $^{(b)}$ 
\\ \hline
0            & 45   & 3.229 & 3.21 &       &       \\
0            & 40   & 3.288 & 3.28 &       &       \\
0            & 37   & 3.336 & 3.33 &       &       \\
0            & 35   & 3.352 & 3.37 &       &       \\
3.10$^4$     & 45   & 2.887 & 2.94 & 4.589 & 4.58  \\
3.10$^4$     & 35   & 2.990 &      & 6.510 & 6.676 \\
\hline
\bat \\
\end{doublespace} \end{table}
%
\begin{table}[H] \begin{doublespace}
\caption{ {\label{tab_d112}} 
Angular dependence of the interaction
between modomain particles. $R$ = 10 $nm$; $K_1(1)$ = $7.10^5$
$J/m^3$ $K_1(2)$ = 0.
$\Theta(\hat{a}_1)$, $\Theta_1$ and $\Theta_2$ denote the angles
($\hat{a_1}$, $\hat{z}$), ($\hat{m}_1$, $\hat{z}$)
and ($\hat{m}_2$, $\hat{z}$) respectively.
$r_{12}$ = 4$R$.
$d_{112}^{(min)}$ is the minimum value of the angular function $d_{112}$
corresponding to $\Theta_1$ fixed and  $d_{112}^{(calc)}$ is the result of the
numeriacal calculation.
$E_{int}$ is the interaction energy per unit volume.
According to the dipolar interaction the theoretical value for
$E_{int}/d_{112}$ is 2072 $J/m^3$.
}
\vskip 1.0 cm
\tab{| *{5}{l} *{1}{c} |}
\hline
$\Theta(\hat{a}_1)$ & $\Theta_1$ & $\Theta_2$ &
$d_{112}^{(calc)}$ & $d_{112}^{(min)}$ & $E_{int}/d_{112}$ ($J/m^3$) \\
\hline
$\pi$/8 & $\pi$/7.948 & $\pi$/4.442 & -1.2019 & -1.202 & 2012 \\
$\pi$/4 & $\pi$/3.987 & $\pi$/2.888 & -1.5828 & -1.584 & 2007 \\
$\pi$/2 & $\pi$/2     & $\pi$/2     & -2.0    & -2.0   & 2004 \\
\hline
\bat \\
\end{doublespace} \end{table}
\saut
\newpage
\subsection*{ Figure captions }
\begin{figure}[h]
\caption{ \label{def_struct_vortex}}
\includegraphics[width = 10.0 cm]{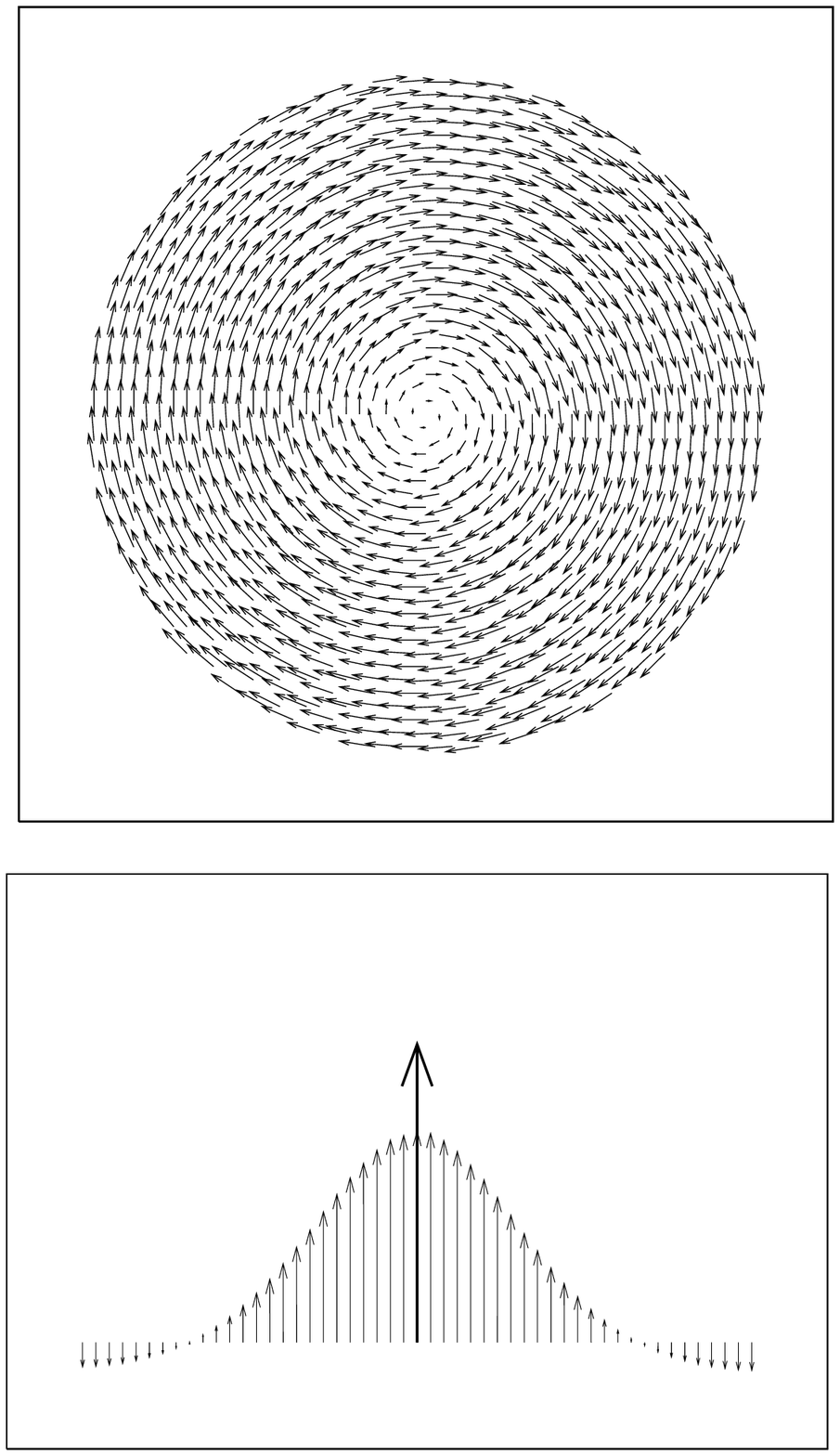}
\end{figure}

\begin{figure}[h] 
\caption{ \label{hyst_r45_k0}} 
\includegraphics[width = 10.0 cm]{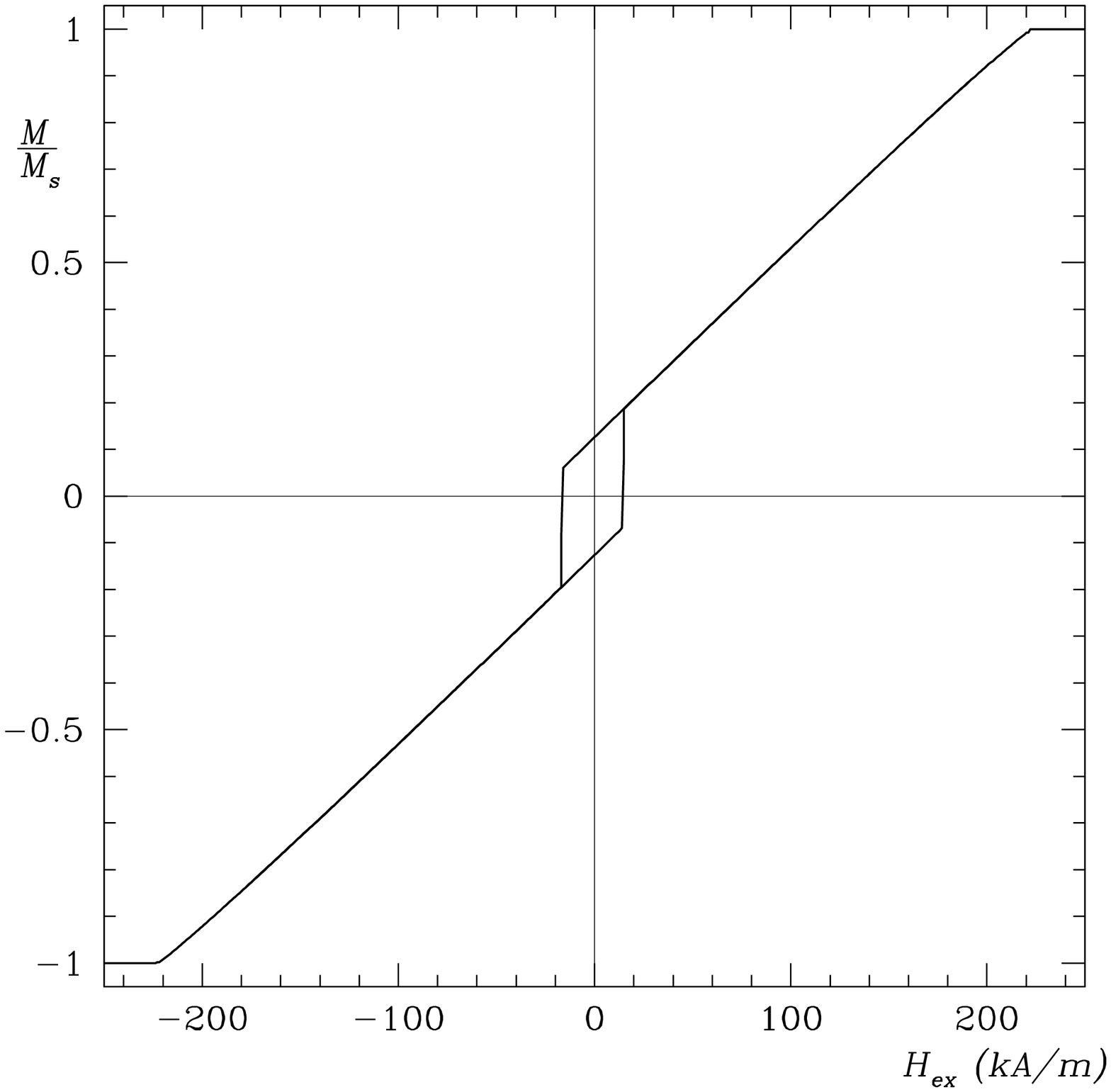}
\end{figure}

\begin{figure}[h] 
\caption{ \label{prof_rem_r45_k0}} 
\includegraphics[width = 10.0 cm]{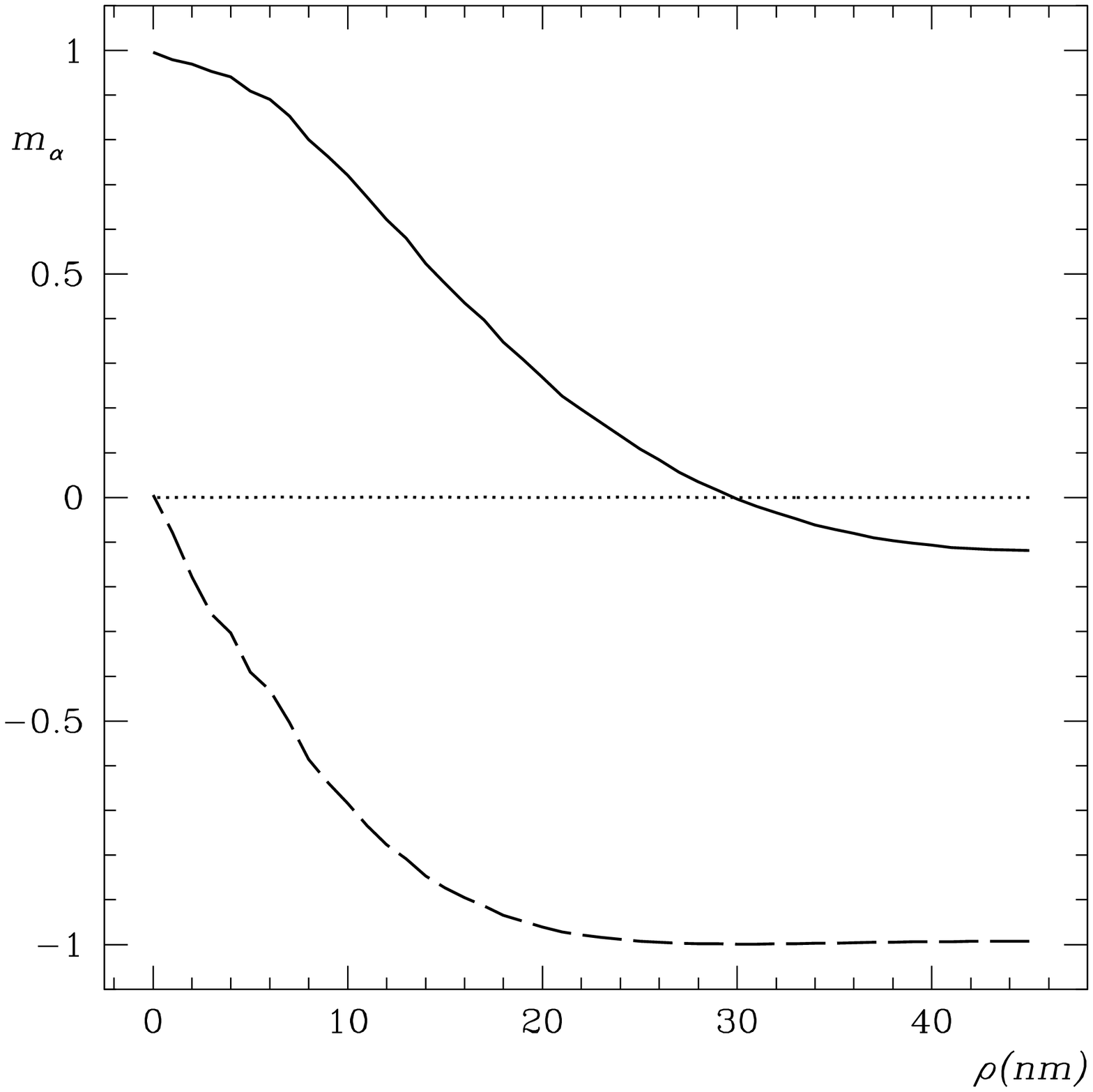}
\end{figure}

\begin{figure}[h] 
\caption{ \label{prof_coer_r45_k0}}
\includegraphics[width = 10.0 cm]{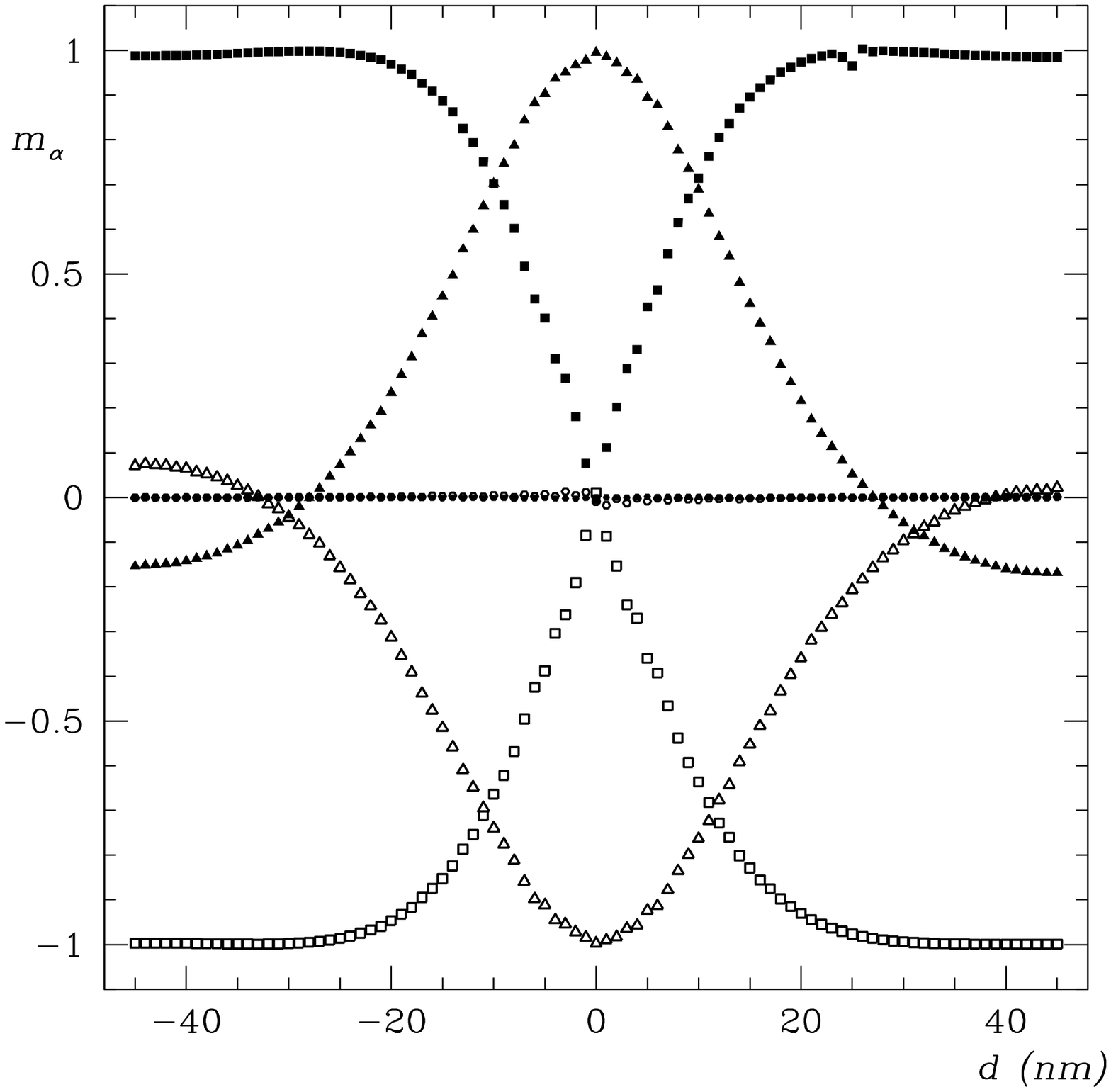}
\end{figure}

\begin{figure}[h] 
\caption{ \label{hyst_r45_k3e4}} 
\includegraphics[width = 10.0 cm]{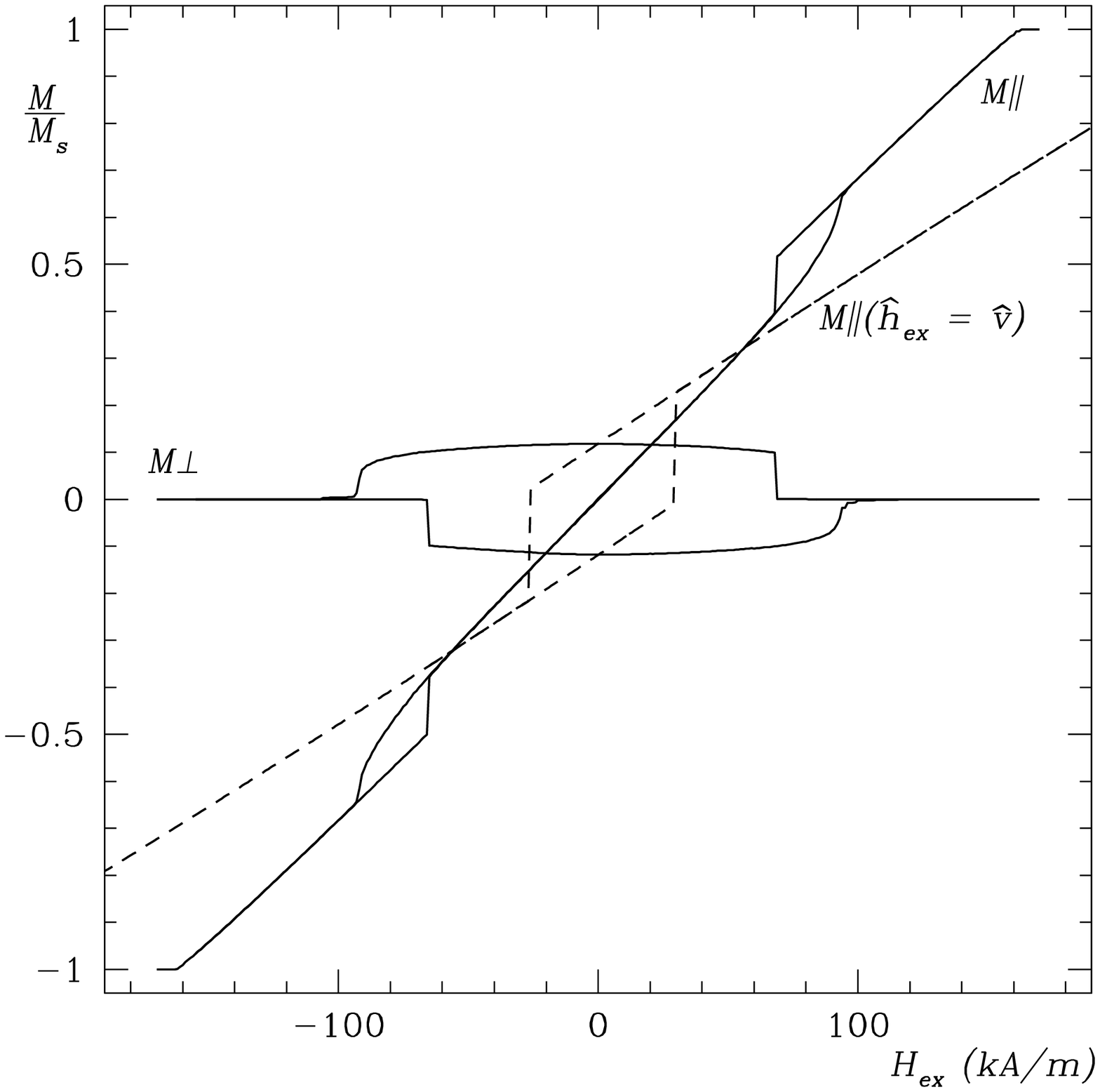}
\end{figure}

\begin{figure}[h] 
\caption{ \label{prof_shift_r45_k3e4}} 
\includegraphics[width = 10.0 cm]{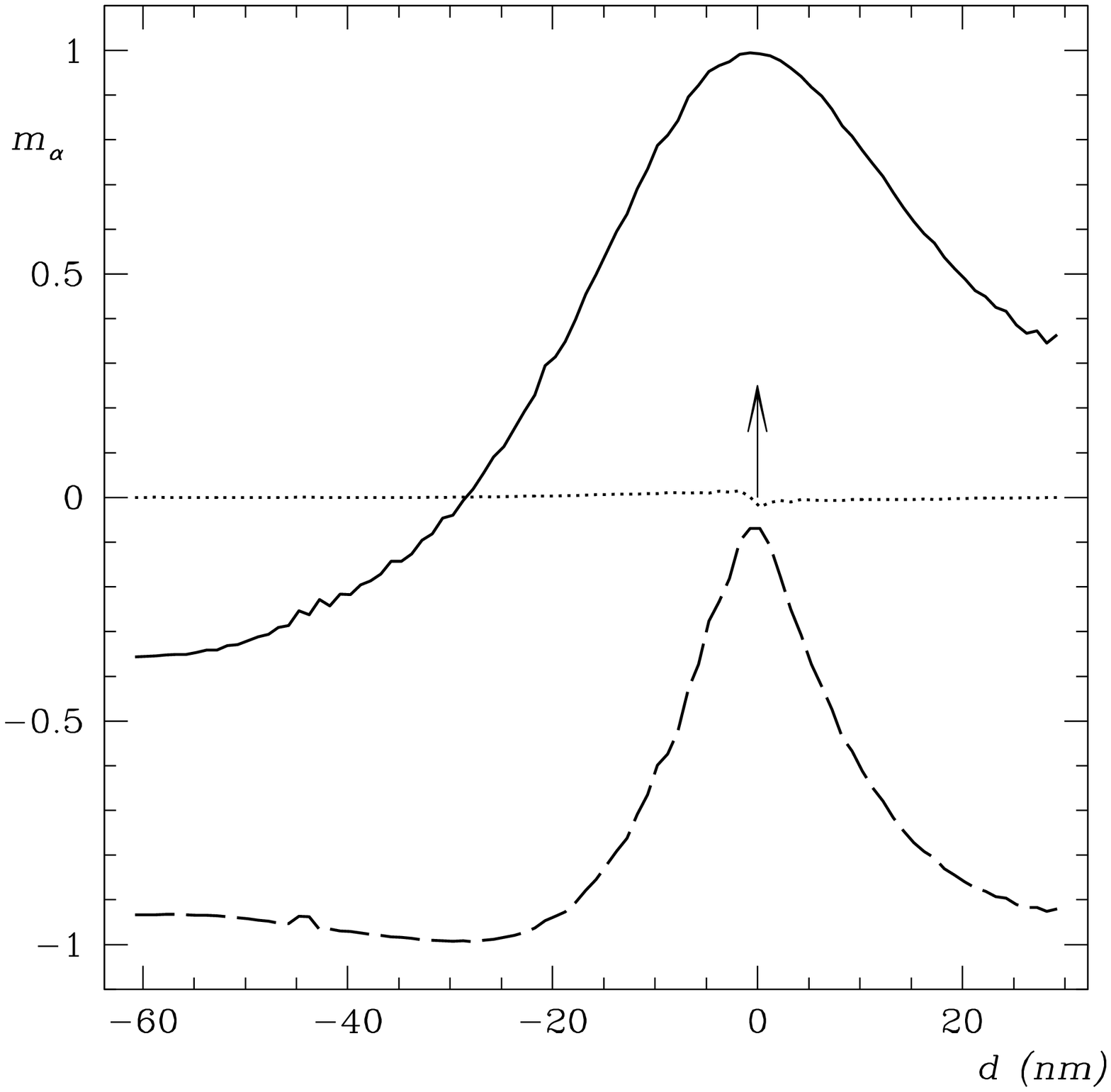}
\end{figure}

\begin{figure}[h] 
\caption{ \label{hyst_taille_k3e4}} 
\includegraphics[width = 10.0 cm]{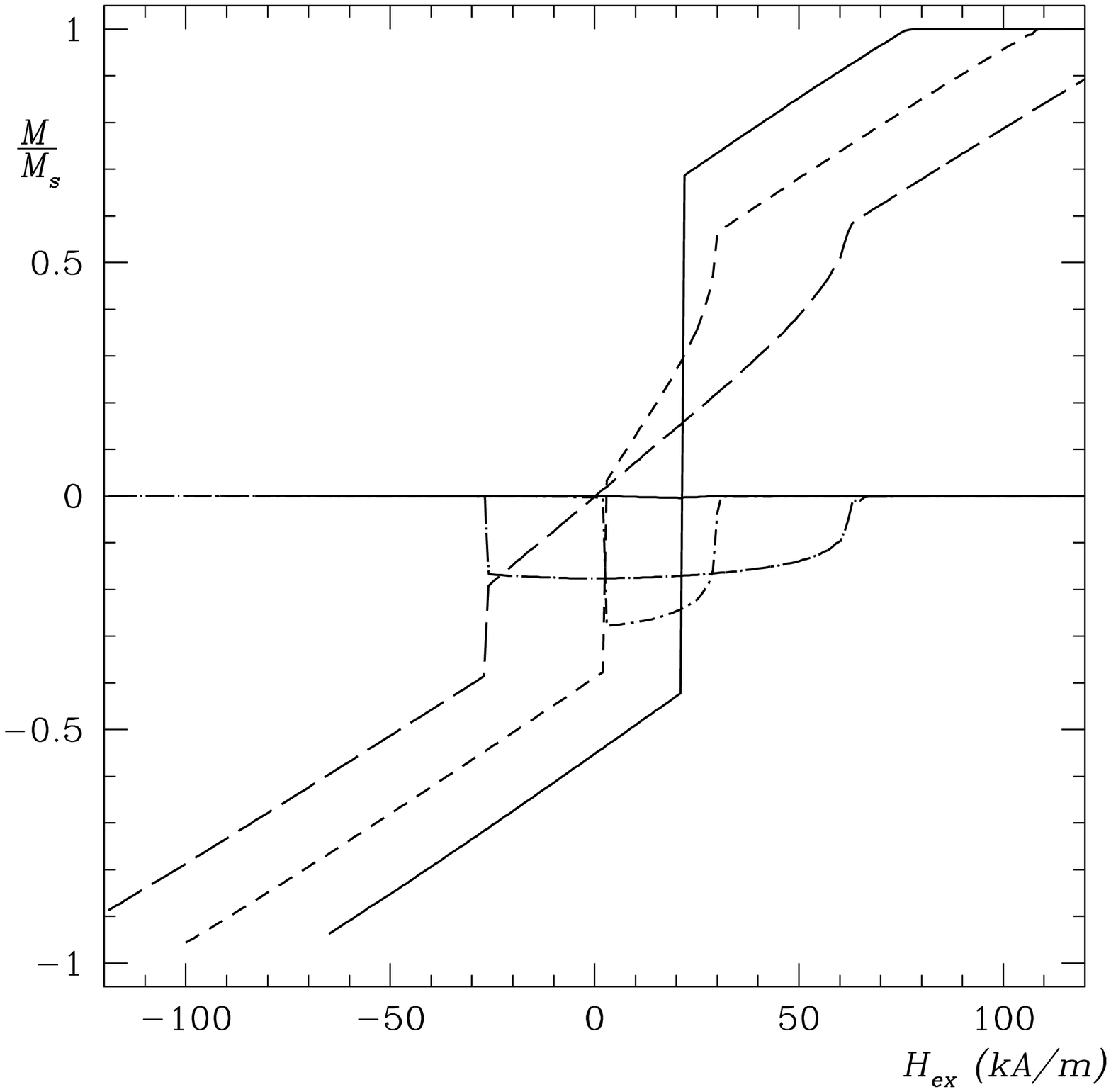}
\end{figure}

\begin{figure}[h] 
\caption{\label{en_rot_r35_k0} }
\includegraphics[width = 10.0 cm]{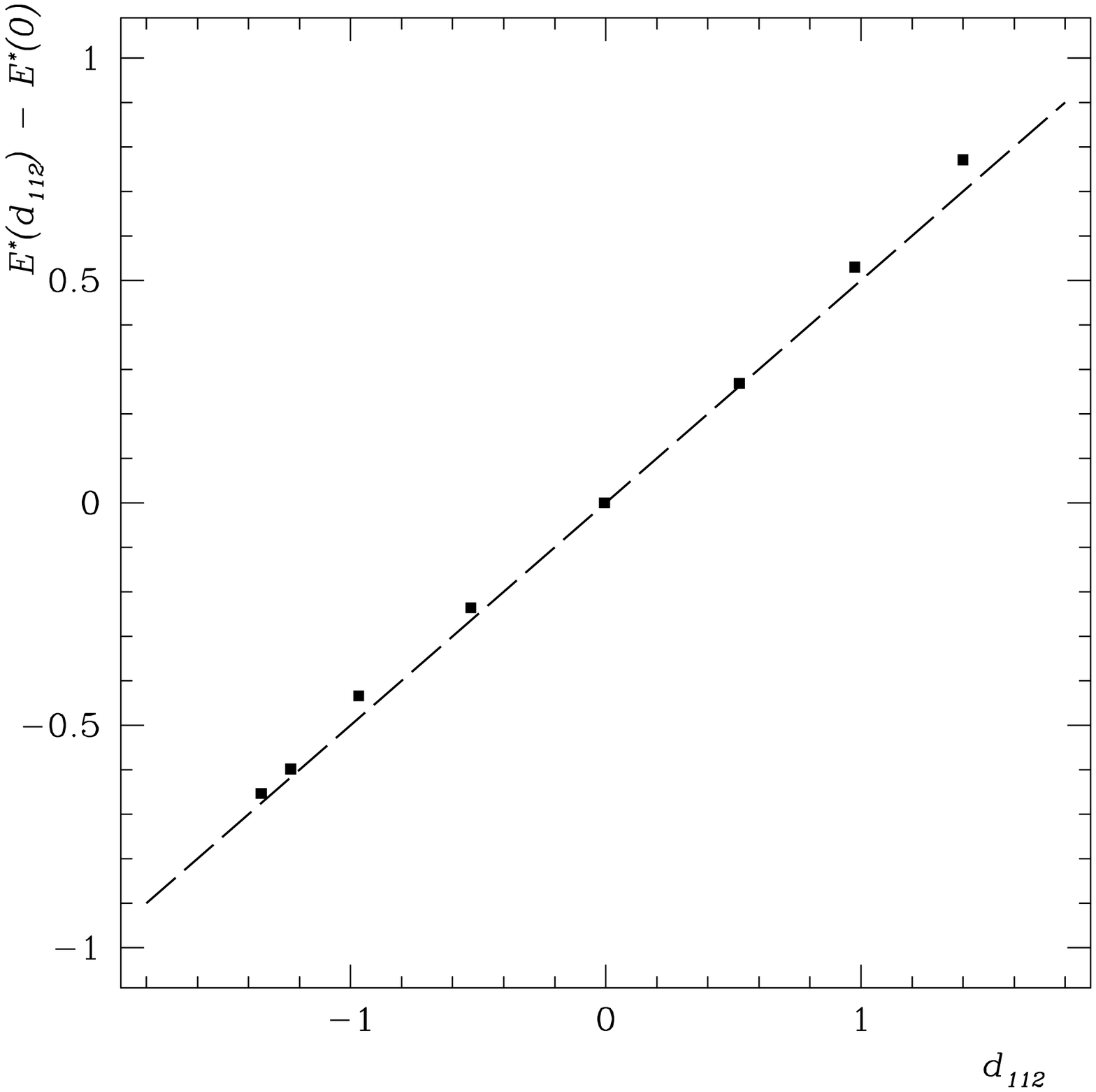}
\end{figure}

\begin{figure}[h] 
\caption{ \label{en_r35_k0_d2} }
\includegraphics[width = 10.0 cm]{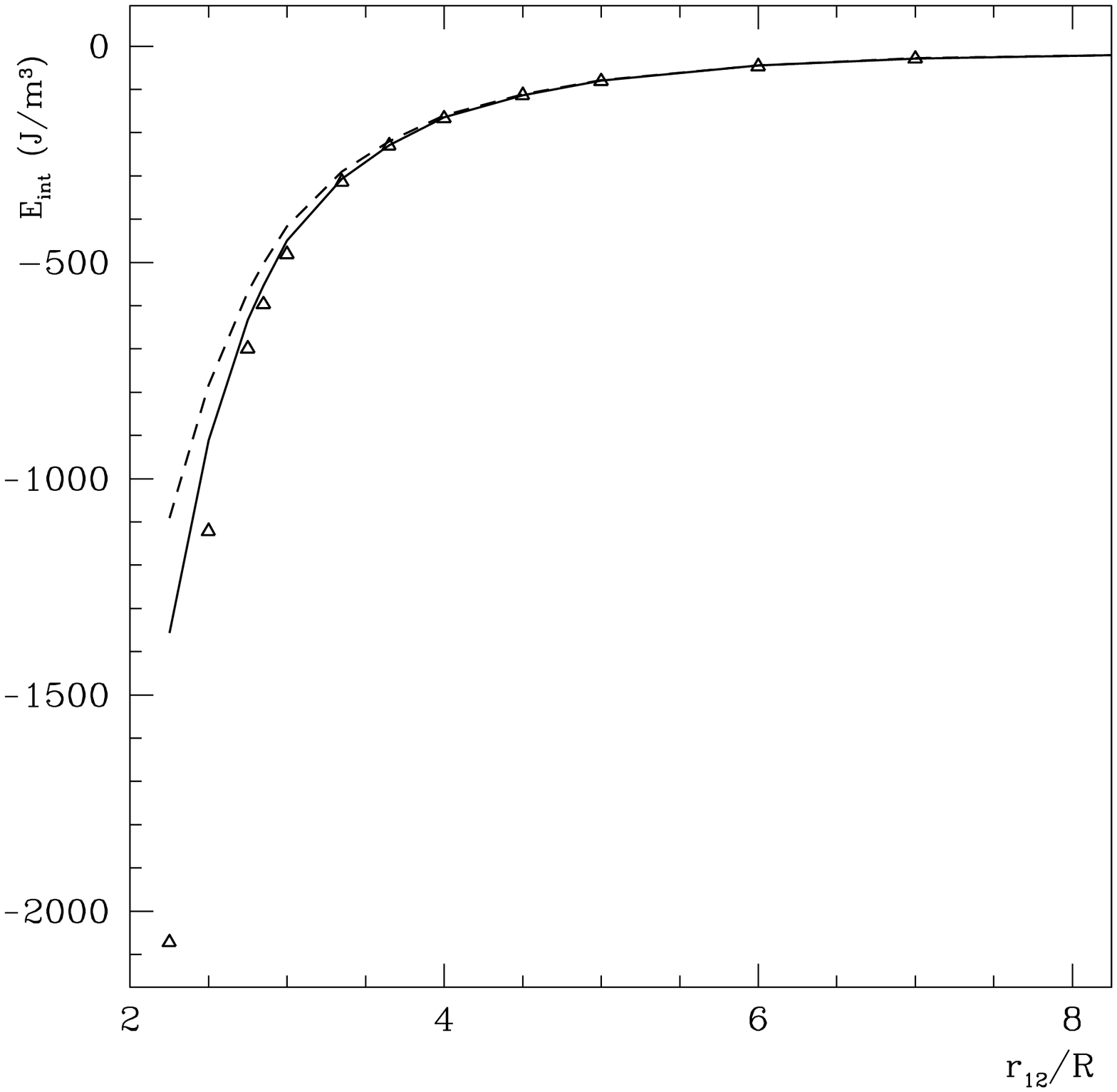}
\end{figure}

\begin{figure}[h] 
\caption{\label{en_tot_ren}}
\includegraphics[width = 10.0 cm]{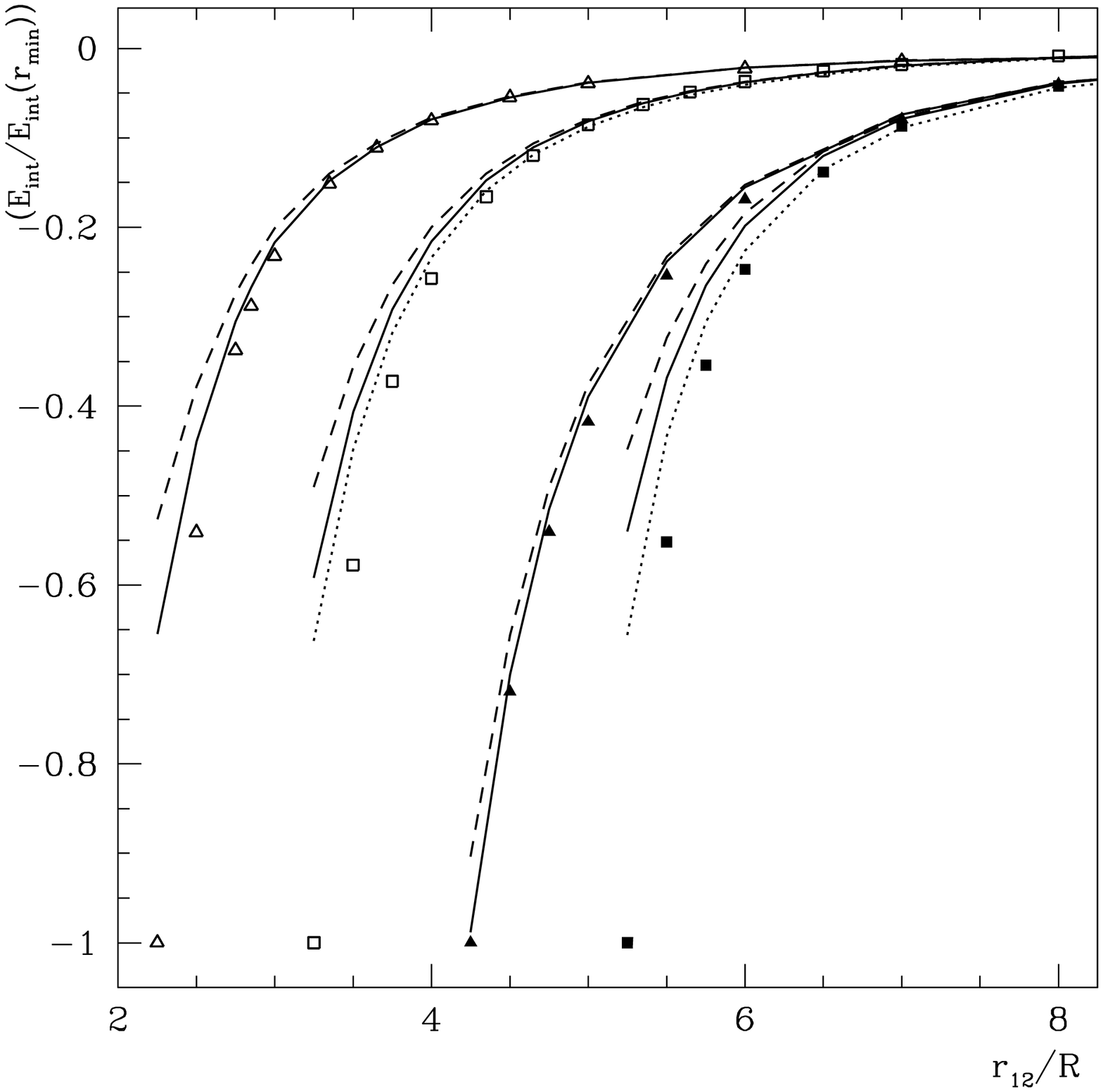}
\end{figure}

\begin{singlespace} 
\begin{itemize} 
\item[Figure \ref{def_struct_vortex}]
Local magnetization structure in the vortex regime. (R = 45 $nm$; $K_1$ = 0;
remanent state.).
Top : projection of the local magnetic moment in the equatorial plane of the
sphere, normal to vortex axis, $\hat{v}$ = $\vec{M}/\Vert{\vec{M}}\Vert$. The length
of the arrows is proportional to the norm of the projection of $\vec{m}(\vec{r})$,
$\vec{m}_p(\vec{r})$. The central part of the vortex is clearly identified as the
region where $\vec{m}_p(\vec{r})$ = 0; the direction of magnetization in this
region coincides with the vortex direction, $\hat{v}$.
Bottom : Local magnetization in the direction normal the equatorial plane of
the vortex shown on top, along a diagonal of this last one. The vortex
direction is shown as the large bold arrow.
\item[Figure \ref{hyst_r45_k0}] 
Magnetization curve in the direction of the field. $R$ = 45 $nm$;
$K_1$ = 0.
\item[Figure \ref{prof_rem_r45_k0}] 
Cylindrical components of the local magnetization profile at remanence. 
across the equatorial plane of the sphere ($z$ = 0).
$m_v$, solid line; $m_{\varphi}$ dashed line; $m_{\rho}$ dotted line.
R = 45 $nm$, $K_1$ = 0.
\item[Figure \ref{prof_coer_r45_k0}] 
Components $m_v$ (triangles), $m_{\varphi}$ (squares) and $m_{\rho}$ (circles)
of the local magnetization profile in the vicinity of the coercive field 
before (solid symbols) and after (open symbols) reversal of the vortex core.
$R$ = 45 $nm$, $K_1$ = 0.
$d$ =  $\rho\; sign(y)$ where $\rho$ is radius in the equatorial plane ($z$ =0).
\item[Figure \ref{hyst_r45_k3e4}] 
Magnetization curve parallel and normal to the field .
$\hat{h}_{ex}$ = $\hat{a}$ (solid line) or $\hat{v}$ (dashed line).
$R = 45nm$, $K_1 = 3.10^4 J/m$.
\item[Figure \ref{prof_shift_r45_k3e4}] 
Components $m_v$ (solid line), $m_{\varphi}$ (dashed line) and $m_{\rho}$ (dotted 
line) of the local magnetization profile relative to the vortex core across the 
($x$ = 0) plane. $R$ = 45 $nm$, $K_1$ = $3\;10^4$ $J/m^3$, $H_{ex}$ = 68 $kA/m$,
$\hat{a}$ = $\hat{h}_{ex}$ = $\hat{z}$ and $\hat{v}$ = $\hat{x}$ .
The vortex is shifted along the $\hat{y}$ axis in the $y\;>\;0$ direction,
by an amount $y_c$ = 15.75 $nm$, leading to
a non symmetric range of variation for $d$. 
The location of the vortex core is indicated by the arrow. 
\item[Figure \ref{hyst_taille_k3e4}] 
Magnetization curve parallel and normal to the field for $K_1$ = $310^4 J/m^3$
and different sizes.
Magnetization in the direction of the field and :
$R$ = 26 $nm$ (solid); 30 $nm$ (short dash); 37 $nm$ (long dash).
Magnetization normal to the field : $R$ = 30 $nm$ (dot short dash);
37 $nm$ (dot long dash).
For $R$ = 26 $nm$, the magnetization normal to the field vanishes and the
magnetization reversal occurs at a positive field since $R$ = 26 $nm$ enters
in the range of particle sizes where the vortex direction is parallel to the
easy axis, chosen as the direction for the field.
\item[Figure \ref{en_rot_r35_k0}] 
Variation of the interaction energy for 2 spheres at $r_{12}$ = 4$R$ with the 
angular function $d_{112}$ characterizing the relative orientations, normalized 
by its maximum value, $E(d_{112} = 2) - E(0)$. $R$ = 35 $nm$; $K_1$ = 0. 
\item[Figure \ref{en_r35_k0_d2}] 
Interaction energy per unit volume between two approaching spheres.
$R=35nm$; $K_1 = 0$; $d_{112} = -2$.
Open triangles: full calculation, from (\ref{delta_e})
(the thin line is a guide to the eye);
solid line: equ.(\ref{inter_ener});
dashed line: simple dipolar approximation, $u = 0$.
\item[Figure \ref{en_tot_ren}]
Same as figure (\ref{en_r35_k0_d2})
for the interaction normalized by the value at $r_{12}$ = 2.25$R$.
dotted lines: result of equ.(\ref{inter_ener})
with the values of $\alpha$ and $u$ fitted
in order to improve the agreement with simulated results.
Open triangles : $K_1$ = 0, $d_{112}$ = -2, $R$ = 35 $nm$;
open squares : $K_1$ = 3 $10^4$ $J/m^3$, $d_{112}$ = -2, $R$ = 35 $nm$;
solid triangles : $K_1$ = 3 $10^4$ $J/m^3$, $d_{112}$ = -1, $R$ = 45 $nm$;
solid squares : $K_1$ = 3 $10^4$ $J/m^3$, $d_{112}$ = -2, $R$ = 45 $nm$.
The different curves are shifted along the $r_{12}$ axis for clarity.
\end{itemize} 
\end{singlespace} 
\newpage
\end{document}